\documentclass{WileyASNA-v1}

\articletype{Proceedings IWARA 2022}%

\received{}
\revised{}
\accepted{}

\begin{document}

\title{Light vector meson photoproduction in ultraperipheral heavy ion collisions at the LHC within the Reggeometric Pomeron approach}

\author[1]{L\'aszl\'o Jenkovszky}

\author[2]{\'Erison S. Rocha}

\author[2]{Magno V. T. Machado*}


\address[1]{\orgdiv{Bogolyubov ITP}, \orgname{National Academy of Sciences of Ukraine}, \orgaddress{\state{Kiev 03143}, \country{Ukraine}}}

\address[2]{\orgdiv{HEP Phenomenology Group}, \orgname{Instituto de Física UFRGS}, \orgaddress{\state{RS}, \country{Brazil}}}


\corres{*M.V.T. Machado. \email{magnus@if.ufrgs.br}}


\abstract{By using the Reggeometric Pomeron model for vector meson production which successfully describes the high energy lepton-nucleon data, we analyse the light meson production in  ultra-peripheral heavy ion collisions at the Large Hadron Collider (LHC). The rapidity distributions for $\rho$ and $\phi$ photoproduction in lead-lead, xenon-xenon and oxygen-oxygen collisions are investigated. }




\maketitle


\section{Introduction}\label{sec1}
The exclusive light vector meson ($V$) photoproduction has been studied in recent years both experimentally and theoretically \citep{H1:2020lzc,CMS:2019awk,ALICE:2020ugp,ALICE:2021jnv,Klein:2020nvu}. The process has  not associated  hard perturbative Quantum Chromodynamics (pQCD) scale in the photoproduction limit, $Q^2\rightarrow 0$. Here, $Q^2$ is the so called photon virtuality in the process $\gamma^*+N\rightarrow V+N$. Therefore, light mesons can be used to test the non-perturbative regime of the strong interactions. Within the  parton  saturation formalism the transition between the region described by pQCD and the non-perturbative regime is interpreted in terms of the 
 nucleon QCD saturation scale \citep{Morreale:2021pnn}, $Q_s(x)\sim x^{-0.3}$, with $x$ being the invariant Bjorken kinematic variable. In the vector meson electroproduction off nucleon target, $x=(M_V^2+Q^2)/(W_{\gamma N}^2+Q^2)$, where $W_{\gamma N}$ is the centre-of-mass energy of the photon-nucleon system.  In the QCD color dipole picture \citep{Nikolaev:1993th,Nemchik:1996pp,Chen:1995pa} $Q_s$ characterizes the boundary on the maximum phase-space gluon density to be reached in the wave-function of the nucleon. In this framework, the light meson photoproduction dynamics at the present accelerator energies can be treated perturbatively as $Q_s$ reaches values $\lesssim 1$ GeV at very high energies. The perturbative description is even improved in case of  scattering off nuclei of atomic number $A$ as the nuclear saturation scale, $Q_{s,A}^2\propto A^{1/3}Q_s^2$, is enhanced regarding the proton. On the other hand, it is well known that this approach has been unable to describe precisely the total photoproduction cross section and $\rho$ production at $Q^2=0$ GeV$^2$ \citep{Goncalves:2020cir,Forshaw:1999uf}. In fact, non-perturbative corrections are necessary and they are embedded in the photon wave-function, $\psi_{T}^{\gamma}$, for color dipoles containing large transverse size.  

In the soft physics sector, the Regge phenomenology  \citep{Jenkovszky:2018itd} is a well founded and appropriated formalism to describe exclusive diffractive processes, including the light meson photoproduction. The vector meson production amplitude is written in a Regge-factorized structure with the corresponding coupling of particles to the Pomeron. The introduction of a perturbative scale dependence suitable for electroproduction can be  constructed based on geometric arguments. The Reggeometric Pomeron (RP) model \citep{Fazio:2013uwa,Fazio:2013hza} is one example of such a class of phenomenological models. The RP model does a good job in describing both photo and electroproduction at the DESY-HERA energy regime considering a nucleon target. The possibility for testing these models in the coherent vector meson production in ultraperipheral heavy ion collisions (UPCs) is a reality nowadays. The basic argument is that the production cross section in nucleus-nucleus ($AA$) collisions can be factorized in terms of the equivalent flux of photons of the colliding nucleus and the photon-target production cross section \citep{Klein:2020fmr}.

In this contribution, the light meson photoproduction in nucleus-nucleus 
 UPCs collisions is investigated. The focus is on the energies and nuclear species in the heavy ion collisions at the Large Hadron Collider (LHC). The theoretical input is the description based on the Reggeometric Pomeron model for the elastic differential and integrated total cross section in the (quasi-real) photon interaction with nucleons. The parameters of the model are consistent with the measurements performed by HERA-H1 \citep{H1:2020lzc} and CMS collaborations \citep{CMS:2019awk} as shown in Ref. \citep{Jenkovszky:2022qnc}. The nuclear coherent cross section is then obtained by using Vector Dominance Model (VDM) and Glauber multiple scattering theory. Predictions are performed for $\rho$ and $\phi$ production in $AA$ UPCs at the LHC. In particular, results are compared to the measurements in PbPb and XeXe UPCs done by ALICE Collaboration \citep{ALICE:2020ugp,ALICE:2021jnv} for the energies of $\sqrt{s_{\mathrm{NN}}}=5.02$ TeV and $\sqrt{s_{\mathrm{NN}}}=5.44$ TeV, respectively. Theoretical estimates for the cross section in OO collisions are also presented.  The study is based on earlier works~\citep{Jenkovszky:2022qnc,Jenkovszky:2021sis,Jenkovszky:2022wcw} by the authors. 

The work has been organized as follows. In Sec. \ref{sec2} we shortly review the exclusive vector meson production, $\gamma+p\rightarrow  V+p$, in the context of the Reggeometric Pomeron model. Afterwards, using VDM model and Glauber formalism for nuclear shadowing, the expression for the coherent nuclear cross section is obtained. In section \ref{sec3} the calculations are compared to available experimental measurements in PbPb and XeXe UPCs collisions at the LHC. Prediction are done for future light ion runs like oxygen-oxygen collisions. Furthermore, discussion on the theoretical uncertainties is presented. In section \ref{sec-con} the key results are summarized.

\begin{center}
\begin{table*}[t]%
\centering
\caption{Values of the parameters for the Reggeometric Pomeron model \citep{Fazio:2013hza}.\label{tab:1}}%
\tabcolsep=0pt%
\begin{tabular*}{40pc}{@{\extracolsep\fill}lcccccccc@{\extracolsep\fill}}
\toprule
\textbf{Meson} & $A_0$ $\left[\frac{\sqrt\text{nb}}{\text{GeV}}\right]$
                 &$\widetilde{Q^2_0}$ $\left[\text{GeV}^2\right]$&   $n$
                 &$\alpha_{0}$& $\alpha'$  $\left[\text{GeV}^{-2}\right]$
                 &$a$&$b$ \\
\midrule
$\rho$ & 344 $\pm$ 376 & 0.29 $\pm$ 0.14      &1.24 $\pm$ 0.07 &1.16 $\pm$ 0.14 & 0.21 $\pm$ 0.53& 0.60 $\pm$ 0.33 & 0.9 $\pm$ 4.3 \\
$\phi$ &  58 $\pm$ 112 & 0.89 $\pm$ 1.40       & 1.30$\pm $ 0.28  & 1.14$\pm$ 0.19 & 0.17 $\pm$ 0.78 & 0.0 $\pm$ 19.8  &  1.34$\pm$ 5.09   \\
\bottomrule
\end{tabular*}
\end{table*}
\end{center}

\section{Theoretical framework}\label{sec2}

Exclusive vector meson photoproduction process, $\gamma +p\rightarrow V+p$, will be described by using a model based on Regge phenomenology, namely the Reggeometric Pomeron model. It is also able to describe electroproduction data as discussed in what follows. In general case, the hardness scale is given by $\widetilde Q^2=Q^2+M_V^2$.   

The elastic differential cross section, $d\sigma_{el}/dt$, related to the single-component Reggeometric model in a given scale $\widetilde Q^2$ is given by \citep{Fazio:2013uwa,Fazio:2013hza}:
\begin{eqnarray}
\label{dsdtel}
\frac{d\sigma_{el}}{dt}& =& \frac{A_0^2\,\exp \left[B_0(\widetilde{Q^2})\,t\right]}{\left(1+\frac{\widetilde{Q^2}}{{Q_0^2}}\right)^{2n}}\left(\frac{W_{\gamma p}^2}{W_0^2}\right)^{2(\alpha(t)-1)}, \\
\label{Bslope}
B_0(\widetilde{Q^2}) & = &  4\left(\frac{a}{\widetilde{Q^2}}+\frac{b}{2m_N^2}\right), 
\end{eqnarray}
where the quantity $B_0(\widetilde{Q^2})$ reflects the geometrical nature of the model and $\alpha (t)$ denotes the effective Pomeron ($I\!\!P$) trajectory. The first and second term in Eq. (\ref{Bslope}) correspond to the effective sizes of the $\gamma I\!\!PV$ and $pI\!\!Pp$ vertices, respectively. In the formula above, $W_0=1$ GeV and $m_N$ is the nucleon mass. It is assumed a linear Pomeron trajectory, $\alpha (t) = \alpha_0+\alpha^{\prime}t$, with an effective Pomeron intercept $\alpha_0$.

Accordingly, the integrated cross section is written as,
\begin{eqnarray}
\sigma(\gamma^*+p\rightarrow V +p) & = &\frac{A_0^2}{\left(1+\frac{\widetilde Q^2}{\widetilde Q^2_0}\right)^{2n}}\frac{\left(W_{\gamma p}/W_0\right)^{4(\alpha_0-1)}}{B\left(W_{\gamma p},\widetilde Q^2 \right)},\\
 B \left(W_{\gamma p},\widetilde Q^2 \right) & = &  B_0(\widetilde{Q^2})+ 4\alpha'\ln \left(\frac{W_{\gamma p}}{W_0}\right).
\label{reggeometric-exp}
\end{eqnarray}
In the photoproduction limit one has $\tilde{Q}^2=M_V^2$ and the parameters of the model for $\rho$ and $\phi$ production are presented in Table \ref{tab:1}. They have been determined \citep{Fazio:2013hza} by using DESY-HERA measurements \citep{H1:1996prv,ZEUS:1997rof,ZEUS:1995bfs,ZEUS:1999ptu,H1:2009cml}.

Now the expressions for the nuclear coherent cross sections are presented. Following the STARLIGHT Monte Carlo generator approach  for UPCs processes \citep{Klein:2016yzr}, nuclear effects for the process, $\gamma +A\rightarrow V+A$ are described here by vector dominance model  \citep{RevModPhys.50.261} and the classical mechanics Glauber  formula for multiple scattering of the vector meson in the
nuclear medium. At $t=0$ the differential cross section  is obtained by using the Optical theorem for scattering in a nucleus and VDM as follows, 
\begin{eqnarray}
\left. \frac{\mathrm{d}\sigma\left( \gamma + A \to V + A \right)}{\mathrm{d}t}\right|_{t=0}
 & = &  \frac{\alpha_{em}}{4 f^2_{V}}\sigma_{tot}^2\left( V A \right), \\
\sigma_{tot}\left( V A \right) & = & \int \mathrm{d}^2 \textbf{b} 
\left[ 1-e^{ -\sigma_{tot}\left( V p \right) T_A\left(\textbf{b} \right) } \right],
\label{glauberVMD}
\end{eqnarray}
where $T_A(b)$ is the nuclear thickness function and $f_{V}$ is the vector-meson coupling. The values $f^2_{\rho}/4\pi = 2.02$ and $f^2_{\phi}/4\pi = 13.7$ are considered in calculations, respectively. For light mesons, $\sigma_{tot}(Vp)$ is large and the cross section $\sigma_{tot}(VA)$ is approximately the  geometric cross section. It is also almost energy independent \citep{Jenkovszky:2022qnc}.

The input for the Glauber model calculation in Eq. (\ref{glauberVMD}) is the effective vector meson–nucleon cross  for the process $V+p\rightarrow V+p $, which is given by:
\begin{eqnarray}
\sigma_{tot}\left( V p \right)= \sqrt{\frac{4f^2_{V}}{ \alpha_{em}} \left.
\frac{\mathrm{d}\sigma\left( \gamma + p \to V + p
  \right)}{\mathrm{d}t}\right|_{t=0}},
  \label{sigrhop}
\end{eqnarray}
where the differential cross section coming from the Reggeometric Pomeron model, Eq. (\ref{dsdtel}), will be introduced in Eq. (\ref{sigrhop}) .
The corresponding integrated cross section is given by:
\begin{eqnarray}
\sigma (\gamma + A \rightarrow V + A) & = &  
\left. \frac{d\sigma (\gamma + A \rightarrow  V + A)}{dt}\right|_{t=0} \nonumber \\
& \times& \int\limits_{t_{min}}^{\infty} \mathrm{d}|t| \, \left| F_A\left(t\right) \right|^2, 
\label{eq:sigtotgammaA}
\end{eqnarray}
where the quantity $F_A$ is the nuclear form factor. It is taken into account an analytic form factor given by a hard sphere of radius, $R_A =r_0A^{1/3}$ fm ($r_0\simeq 1.2$ fm), convoluted with a Yukawa potential with range $a$ \citep{PhysRevC.14.1977},
\begin{eqnarray}
 F_A(|q|) & = &  \frac{4\pi\rho_0}{A |q^3|} \left( \frac{1}{1+a^2q^2} \right) \nonumber \\
 &\times & \left[ \sin{(|q| R_A)} - |q| R_A\cos{(|q| R_A)}  \right], 
\end{eqnarray}
where $q$ is the momentum transfer, $\rho_0 = 3A/(4\pi R_A^3)$ fm$^{-3}$ and $a = 0.7$ fm.

In the calculations the reggeon contribution is added to the photoproduction off nucleons. The corresponding cross section is parameterized as,
\begin{eqnarray}
\left. \frac{d\sigma^{I\!\!R} (\gamma p\rightarrow VP)}{dt}\right|_{t=0} = b_VYW_{\gamma p}^{-\eta}, 
\end{eqnarray}
where the constants $b_V= 11$ GeV$^{-2}$, $Y = 26.0$ $\mu$b and $\eta =  1.23$ have been considered for the $\rho$ production. For the $\phi$, meson exchange is strongly suppressed, and the reaction occurs only through $I\!\!P$-exchange. 

\begin{figure}[t]
\centerline{\includegraphics[width=.55\textwidth]{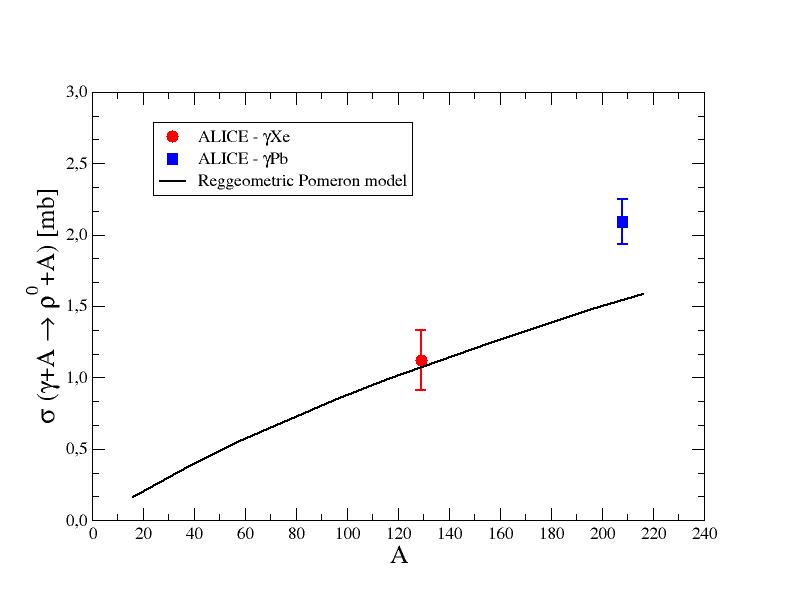}}
\caption{The $A$-dependence of the  cross section for the coherent production of $\rho$ meson from Reggeometric Pomeron model at the LHC. Data from ALICE Collaboration \citep{ALICE:2021jnv}. \label{fig:1}}
\end{figure}

The $A$-dependence of the  cross section for the coherent production of $\rho$ meson, $\sigma (\gamma +A \rightarrow \rho + A)$, is presented in Fig. ~\ref{fig:1}. Comparison is done with the extracted values of the coherent cross section performed in  Ref. \citep{ALICE:2021jnv} by ALICE Collaboration  using the measured data on UPCs at the LHC (PbPb collisions at 5.02 TeV and XeXe at 5.44 TeV). The description is quite reasonable for the nuclear dependence. At central rapidity, $y=0$, the photon-nucleon centre-of-mass energy squared is $W_{\gamma N}^2 = M_V\sqrt{s_{\mathrm{NN}}}$. For xenon the cross section corresponds to $W_{\gamma N}=65$ GeV and $\sigma (\gamma \mathrm{Xe}\rightarrow \rho \mathrm{Xe}) \simeq 1.12 \pm 0.21$ mb. The predicted values from the Reggeometric Pomeron model is 1.07 mb. For lead, the data is $\sigma (\gamma \mathrm{Pb}\rightarrow \rho \mathrm{Pb}) \simeq  2.09\pm 0.16$ mb  for energy $W_{\gamma N}= 62$ GeV and the prediction  1.54 mb. Theoretical calculation underestimates the extracted $\gamma$Pb cross section, which suggests a strong nuclear shadowing correction for very large nucleus in the formalism considered for the study.

\begin{figure*}[t]
\centerline{\includegraphics[width=.75\textwidth]{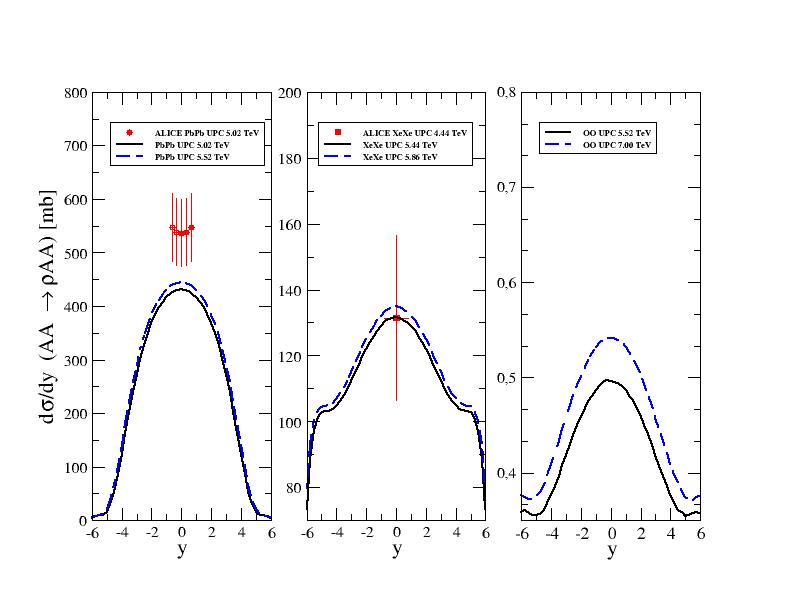}}
\caption{Rapidity distributions for the exclusive $\rho$ meson photoproduction in ultraperipheral PbPb (left panel), XeXe (central panel  and OO (right panel) collisions considering the Reggeometric Pomeron model. Prediction are done for the current run (solid lines) and future HL-LHC run (dashed lines). Comparison is done to 
 ALICE Collaboration data \citep{ALICE:2020ugp,ALICE:2021jnv}. \label{fig:2}}
\end{figure*}
\section{Results and discussions}\label{sec3}

The rapidity distribution for meson production in nucleus-nucleus UPCs takes a factorized form in the Equivalent Photon Approximation (EPA). The expression is given by:
\begin{eqnarray}
\frac{d\sigma (A+A\rightarrow A+V+A)}{dy} & = & k^+\frac{dN_{\gamma/A}(k^+)}{dk}\sigma_{\gamma A\rightarrow VA}(k^+)\nonumber \\
&+ & k^-\frac{dN_{\gamma/A}(k^-)}{dk}\sigma_{\gamma A\rightarrow VA}(k^-),\nonumber \\
\end{eqnarray}
where $dN_{\gamma/A}/dk$ is the photon flux in nucleus $A$ and $k$ is the photon momentum. For fixed rapidity $y$ and transverse momentum $p_T^2 \approx |t|$ of the produced mesons, the photon momentum is given by $k^{\pm} = \frac{M_{V}^2-t}{2M_Te^{\mp y}}$. Here, $M_T=\sqrt{M_V^2+p_T^2}$ is the transverse mass of the mesons. 

For simplicity, the analytical expression for the flux of photons produced by a fast-moving point-like charge has been considered \citep{Klein:2016yzr}, 
\begin{eqnarray}
\frac{dN_{\gamma/A}(k)}{dk} =\frac{2Z^2\alpha_{em}}{\pi k}\left[xK_0(x)K_1(x) - \frac{x^2}{2}  \left(K_1^2(x)-K_0^2(x)  \right) \right], \nonumber \\
\end{eqnarray} 
where $x= 2R_Ak/\gamma_L$ and $\gamma_L$ is the Lorentz factor. $K_{0,1}(x)$ are the modified Bessel functions of the second kind.

\begin{figure*}[t]
\centerline{\includegraphics[width=.75\textwidth]{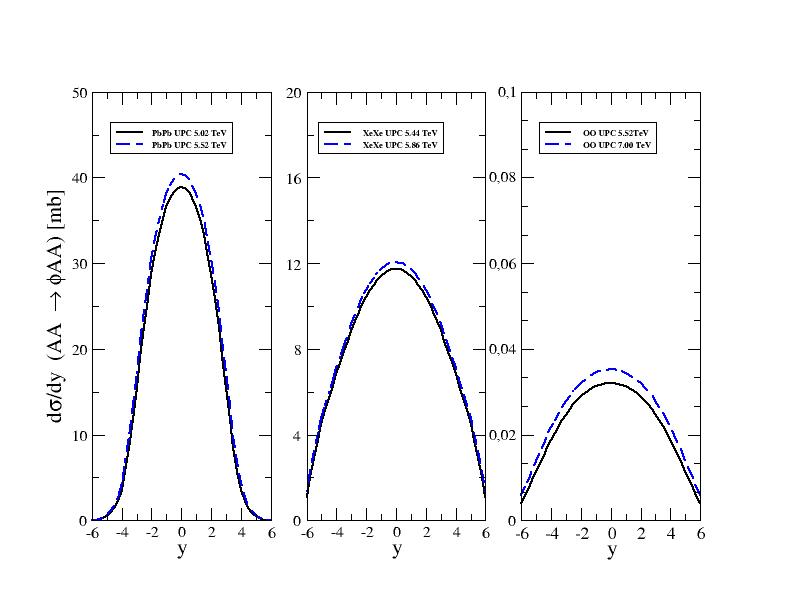}}
\caption{Rapidity distributions for the exclusive $\phi$ meson photoproduction in ultraperipheral PbPb (left panel), XeXe (central panel and OO (right panel) collisions considering the Reggeometric Pomeron model. Prediction are done for the current run (solid lines) and future HL-LHC run (dashed lines). \label{fig:3}}
\end{figure*}

In Fig. \ref{fig:2} results are shown for $\rho$ production in PbPb, XeXe and OO UPCs at the LHC in the rapidity range $|y|\leq 6$. Left panel: predictions are presented for the PbPb collisions in energies of $\sqrt{s_{\mathrm{NN}}} = 5.02$ (solid line) and   5.52 TeV (dashed line), respectively. It is shown also the measurement performed by ALICE Collaboration  at mid-rapidity \citep{ALICE:2020ugp}. Central panel: predictions for XeXe collisions in $\sqrt{s_{\mathrm{NN}}} = 5.44$ (solid line) and  
 5.86 TeV (dashed line) compared to ALICE data \citep{ALICE:2021jnv}. Right panel: predictions for OO collisions with energies of  $\sqrt{s_{\mathrm{NN}}} = 5.52$ (solid line) and  
 7.00 TeV (dashed line), respectively. The second energy bin corresponds to the designed $\sqrt{s_{\mathrm{NN}}}$ for the future  High-Luminosity LHC (HL-LHC) run  \citep{Bruce:2018yzs}.   In general, the model is suitable to predict the magnitude and shape of the rapidity distribution in XeXe UPCs. The corresponding suppression at central rapidities in PbPb case is consequence of the coherent cross section to be underestimated as shown in Fig. \ref{fig:1}  Namely, the nuclear effects for xenon nucleus are less intense as for lead nucleus. 

 The contribution of Reggeons to $\rho$ coherent production turns out to be evident in the rapidity distributions at large $|y|$. It is also $A$-dependent where the contribution is more important for light nuclei than heavy ones. In particular, the energy dependence of the photon-nucleon cross section, the suppression due to nuclear shadowing, and the drop of the flux of high-energy photons drive the distribution in the central and forward (backward) rapidity regions. Bumps or shoulders at large rapidities are due to an enhanced contribution of low-energy
photoproduction related to the secondary Reggeon exchange in the meson-nucleon interaction. The Glauber shadowing at
low energies is more intense for lead nuclei compared to xenon and oxigen ones. This is the reason for the shoulder appearing in XeXe and OO collisions and not in PbPb.

Finally, in Fig. \ref{fig:3} predictions for coherent $\phi$ photoproduction are presented. Using the same notation as previous figure, calculations are performed for PbPb, XeXe and OO collisions for the energies of the present LHC run (solid lines) and the HL-LHC run (dashed lines). Currently, there is no data available for $\phi$ production in AA UPCs at the LHC. It is planned a high-granularity  detector named FoCal \citep{Bylinkin:2022wkm} to be installed at the ALICE experiment, covering large rapidities.  It  will allow to measure the cross sections and expected yields for exclusive production in the dielectron decay channel with 
 a coverage for both electrons within $3.4 \leq \eta \leq 5.8$. The detector FoCal can contribute for precise measurements of low-mass vector mesons production  such as $\rho$ and $\phi$ as well as  excited $\rho$ meson states.

 Finally, we discuss the theoretical uncertainties on the calculations. The predictions are in agreement with those from the STARLIGHT Monte Carlo generator for UPCs \citep{Klein:2016yzr} but the cross sections of the coherent meson production are considerably smaller than calculations in Refs. \citep{Frankfurt:2015cwa,Guzey:2020pkq}. The main sources of discrepancies are the use of the factorized form, Eq. (\ref{eq:sigtotgammaA}), and the classical Glauber formula, Eq. (\ref{glauberVMD}). It is considered  the inelastic meson–nucleus cross section instead of the total cross section which decreases the prediction for the forward cross section by a factor $\sim 2$. Namely,  the total cross section of the $VA$ interaction is obtained from classical mechanics (MC) Glauber model. However, the quantum mechanics expression is given by the Gribov-Glauber (GG) formalism where the $VA$ cross section is given by:
\begin{eqnarray}
\sigma_{tot}^{\mathrm{GG}}(VA) = 2\int d^2\vec{b} \left[ 1-\exp \left( -\frac{1}{2}\sigma_{VN}T_A(\vec{b})\right) \right].
\end{eqnarray}
For example, in the simplification of a sharp sphere nucleus with $\rho_0=0.17$ fm$^{-3}$ and radius $R_A$ one can obtain an estimate of the ratio between the GG and CM cross sections, 
\begin{eqnarray}
\frac{\sigma_{tot}^{\mathrm{GG}}(VA)}{\sigma_{tot}^{\mathrm{CM}}(VA) }\approx 2\left(1-\frac{3}{2\rho_0^2\sigma_{VN}^2R_A^2}  \right) .
\end{eqnarray}
Let us consider the $\rho$ production. The ratio is $\approx 1.67$ for lead  and $1.55$ for xenon by using $\sigma_{\rho N}\approx 25$ mb. The classical probabilistic formula (CM) and the Glauber-Gribov (GG) approach give near values of the $\sigma_{tot}(VA)$ only when $\sigma_{tot}(Vp) T_A(b) \ll 1$. It is expected that difference for $\phi$ production be smaller due to the lower $\sigma_{tot}(\phi A)$ cross section.

\section{Conclusions}\label{sec-con}
In this contribution predictions for exclusive light vector meson photoproduction in UPCs collisions at the LHC are presented and compared with the current experimental measurements. The theoretical approach is based on Regge phenomenology. In particular, the single-component Reggeometric Pomeron model has been considered. Concerning the rapidity distributions for $\rho$ production in PbPb UPCs, the model underestimates the 
 data whereas does a better job in case of XeXe UPCs. Predictions are provided for OO UPCs in a future LHC run in light heavy ion model.  
 The results for $\phi$ follow the same trend observed in $\rho$ production. The theoretical uncertainties are considerably large concerning the computation of nuclear effects, factorization between energy and momentum transfer dependence among others.

The main focus was on the  investigation of how models of vector meson
production in electron-proton scattering affect the results in
ultra-peripheral nucleus-nucleus collisions. This direction of research is especially promising also because of the planned experiments
at future accelerators. It is promising the coverage of the ALICE FoCal detector which allows to study the vector meson photoproduction in both low and high photon-nucleon centre-of-mass energies.


\section*{Acknowledgments}
 This work was partially supported by the \fundingAgency{National Academy of Science of Ukraine} grant \fundingNumber{1230/22-1 Fundamental Properties of Matter}, the \fundingAgency{Coordination for the Improvement of Higher Education Personnel (CAPES/Brazil)} grant \fundingNumber{Finance Code 001} and by the \fundingAgency{National Council for Scientific and Technological Development (CNPq/Brazil)} grant \fundingNumber{306101/2018-1}.



\subsection*{Financial disclosure}

None reported.

\subsection*{Conflict of interest}

The authors declare no potential conflict of interests.



\noindent

\bibliography{JRM-IWARA22.bib}%



\end{document}